\documentclass[guide]{rmaa} 
\title{Mass determination for T Tauri stars from JHK Photometry}
\author{Tania A. L\'opez-Chico \& Luis Salas}
\fulladdresses{
\item T. L\'opez-Chico and L. Salas: Instituto de Astronom\'{\i}a, Universidad Nacional Aut\'onoma de M\'exico, Apdo. postal 877, 22830 Ensenada, B.C., M\'exico (tania@astrosen.unam.mx), (salas@astrosen.unam.mx)} 
\shortauthor{L\'opez-Chico \& Salas} \shorttitle{Mass determination  for T Tauri stars from JHK Photometry} 
\ReceivedDate{na} 
\AcceptedDate{na} 
\SetYear{2007}
\resumen{Presentamos un m\'etodo que utiliza modelos de discos para determinar masas de estrellas T Tauri a partir de fotometr\'ia JHK.\@ Encontramos que el exceso infrarrojo producido por el disco se comporta de forma similar a un vector de extinci\'on cuando son graficados en diagramas color-color y color-magnitud, pero son linealmente independientes.\@ Mediante la manupulaci\'on de estos vectores como si fueran una base de un espacio vectorial, llevamos a cabo una transformaci\'on de coordenadas que nos permiti\'o encontrar la masa de la estrella central para una edad dada.\@ Para probar este m\'etodo basado en vectores principales (PV), comparamos los valores de masa obtenidos ($M_{PV}$) en 14 objetos T Tauri de la regi\'on de Taurus-Auriga con los valores de masa calculados por m\'etodos din\'amicos.\@ Tambien analizamos 4 sistemas m\'ultiples   (GG Tau, FO Tau, FS Tau y V733 Tau) y determinamos la masa de cada componente del sistema.  Se analiz\'o el efecto que se produce al utilizar diferentes modelos evolutivos y la variaci\'on de otros par\'ametros en la determinaci\'on de masas este m\'etodo. } 

\abstract{We present a method that uses disk models to determine stellar masses for Pre-Main sequence T Tauri stars from JHK photometry.\@ We find that the infrared excess produced by the disk behaves in a similar way to the extinction vector when plotted on color-color and color-magnitude diagrams, but that it is linearly independent.\@ 
Employing these vectors as a basis of a vector space we carry out a coordinate transformation that allows us to find the mass of a central star for a given age.\@ To test this Principal Vectors (PV) method we compare the mass values ($M_{PV}$) obtained for 14 T Tauri objects in the Taurus-Auriga region with mass values known by dynamical methods.\@ Further on, we analyzed 4 systems with multiple components (GG Tau, FO Tau, FS Tau, and V733 Tau) and determined the mass for each component of the system. We analyzed the effect of using different evolutionary models and other  parameter values.}

\makeglossary{} 
\begin{document}
\maketitle
\tableofcontents 
\section{Introduction}
\label{sec:introduction}

The only direct method to determine stellar masses is through the analysis of dynamical parameters.  This can be accomplished by measuring periods of orbiting components, radial velocities of unresolved companions, or from the kinematics of a circumstellar disk. 
Obtaining the necessary data is often impossible, or else difficult, and so dynamical methods have been used only for a few pre-main sequence (PMS) stellar objects.
As a consequence of this, most of the mass values are obtained by secondary methods.\@

The usual secondary method is to place the objects on a Temperature-Luminosity diagram along with evolutionary models (\citealt{coh79}).  Many circumstances encumber this task. First, temperature could be characterized from the colors of the objects.  But colors are affected by circumstellar disks.  The disks obscure optical observations and produce infrared (IR) excess, also the accretion onto the surface of the star increases the UV part of the spectrum and veils the optical, and a reddening is produced by extinction from the dusty medium.\@

Therefore, temperature is better determined from the spectral type of the object, obtained through spectroscopy (\citealt{p2}; \citealt{p32}; \citealt{p62}; \citealt{p113}; \citealt{p131}; \citealt{p171}; \citealt{p175}; \citealt{p193}; \citealt{p194}; \citealt{p211}).  
To determine the luminosity of the source, extinction corrections have to be applied.
These corrections can also be determined from the spectral type of the source  by comparing the observed colors with the intrinsic ones. In doing this task, the use of colors where either the UV excess due to accretion or the IR excess due to disk emission are usually avoided because both effects contaminate the colors.  Thus, colors involving the bandpases R, I and J are preferred (\citealt{p113}; \citealt{p171}).  Among several  improvements, \citet{har91} have refined the method by modeling the veiling to correct the spectrum and so, by this way, he obtains a better estimate of the spectral type and the extinction corrections. However, the acquisition of spectroscopic data is a difficult task for faint objects in dusty environments, which constitute the vast majority of PMS stellar objects. This has lead many authors to explore the initial mass function (IMF) of clusters in the IR.

The most used measure of  the IMF is the K luminosity function (KLF) because it is likely to produce a faithful representation of the true IMF of a cluster (\citealt{p139}).  This is so because the IR excess is only a small correction when large KLF bins are employed.\@   
However, the K luminosities of objects in clusters are affected by extinction.  Although this problem sometimes is not accounted for, most authors either correct for extinction using the color-color diagram (C-C D) (\citealt{p1}; \citealt{p3}; \citealt{p5}; \citealt{p132}; \citealt{p161}; \citealt{p171}; \citealt{por00}) or the color-magnitude diagram  (C-M D) (\citealt{p1}; \citealt{por00}).  On the other hand, the K luminosity is seldomly corrected by the IR excess emission (however see \citealt{p132}) and authors that worry about it, sometimes favor the use of the JLF or HLF where the contamination is smaller (\citealt{p109}; \citealt{p111}; \citealt{p190}; \citealt{p192}). 

But  the IR excess emission has been clearly identified and shown to have a definite effect in C-C and C-M diagrams.\@ \citet{lad92} and \citet{hil92} show that objects of different evolutionary states occupy definite regions of the C-C D (see also \citealt{pal93}).\@ The classical T Tauri star locus \citep{mey97} is a well defined linear region in the C-C D. This locus is occupied by dereddened T Tauri stars from a Taurus sample.\@ 
In their work,  \citep{mey97} displace T Tauri stars along the reddening vector by an amount determined from the E(R$_c$-I$_c$) excess above the intrinsic colors known from the spectral type of each source.\@ The resulting locus lies along a narrow line redwards of the main sequence colors corresponding to stars of spectral types around M.\@ That is,  the excess in the colors, in particular  E(J-H) and E(H-K), are related to each other and define an excess vector in the C-C D.\@  Then, \citet{mey97} model the excess emission from accretion disks around K5 to M5 stars to show that this locus is indeed well explained and expected due to the existence of circumstellar disks.\@ 
Therefore, it is proven that by adding an excess vector and an interstellar extinction vector to a stellar photosphere, one can recover the position of a particular T Tauri star on the C-C D.\@  However, if the spectral type of the star is not known, the magnitudes of the vectors can not be assessed. \@ 

The existence of a vector like excess, this time for the K vs H-K  diagram, has been found by \citet{hc2000}.\@  They analyzed a similar sample of T Tauri stars in Taurus and found a linear relationship between the K excess emission and the excess in H-K.  Again, the excesses are determined from the dereddened colors and magnitudes  if the intrinsic colors are known from the spectral types of the sources.\@ In their work, \citet{hc2000} go on to obtain the IMF of the trapezium cluster by using the vector sum of this excess vector and the interstellar extinction vector.\@ However, since they only used the C-M D, they had incomplete information to obtain the magnitudes of these two vectors, so they relied on a statistical approach that treats these unknown quantities as probability distributions. 

Therefore, it has been shown that IR excess emission affects both near infrared (NIR) C-C and C-M diagrams in a similar way as the interstellar extinction.\@ However, the two diagrams have not been used simultaneously to obtain the magnitude values of IR excess and extinction.  In this work we will explore this possibility.\@ We will begin by analyzing the effect of a circumstellar disk on the NIR colors and magnitudes. Theoretical models to explain disks may turn out to be very complicated and include many parameters \citep{joh01}.\@  The first circumstellar disk models were assumed geometrically thin and optically thick  \citep{sha73}.\@  \citet{ken87} realized that the thickness of the disk should increase radially, leading to a flared disk.\@  This resulted ia a better fit to the spectral energy distribution (SED) for many T Tauri stars.\@  This disk model assumed that the temperature was only dependent on the disk radius.\@ The vertical temperature gradients in the disk were first taken into account by \citet{cal91,cal92}, with further analysis by \citet{dal98,dal99}, and \citet{chi97,chi99}.\@  In these models the absorption of stellar radiation occurs only in the top disk layers, so the heating of the internal disk is produced by viscous dissipation.\@  \citet{dal98,dal99} calculated the spectral emission of a disk assuming the same structure for the gas and dust in the disk, while \citet{chi99}, ignored the gas contribution.\@ \citet{dal98} show that if there is a low coupling between the gas and dust at the surface of the disk, the gas temperature decreases.\@ In that case, the dust dominates the disk emission in classical T Tauri stars observations.\@

\citet{dul01} extended the \citet{chi97} model to include a puffed-up inner rim for Ae/Be stars. \citet{muz03} showed that the emission of a puffed rim also fits the  SED of T-Tauri stars in the 2-5$\mu$m region.  
The rim models were rather simplified treating them as a vertical
walls ( as in models by \citealt{dal05}).  Rounded-off rims seem to be more consistent with observations \citep{ise05}.

We will explore the effects of the disk excess emission in C-C and C-M Diagrams using reprocessing models and more complicated accretion models.\@ We will show that in certain cases the produced excess can be interpreted as a reddening vector.\@ Then, we will show on the diagrams that these vectors point out in a different direction than the interstellar extinction one, and so we will identify a pair of Principal Vectors (PV's) produced by the disk ( $\vec{D}$) and interstellar extinction ($\vec{X}$).\@  These vectors can be used as a tool to determine stellar masses from JHK photometry, without the need of spectral types.\@

A set of PMS tracks at the cluster's age will be needed to read out the mass of the star.\@ 
We will take advantage of recent work on evolutionary models like those of  \citet{dan94,dan97} (hereafter DM97) ;\citet{pal99}(hereafter PS99); \citet{sie00} (hereafter S00) ,\citet{bar98} (hereafter BCAH98) and the Yale Group $Y^{2}$ \citet{yi01} (hereafter $Y^{2}$01).\@ \citet{hil04} reviewed these models listing the basic parameters employed by each one.

In \S 2 we proceed with the calculation of the NIR excess emission of disks from reprocessing and accretion models and obtain the PV for each diagram.\@  Also,  we will describe the procedure to obtain mass values with the PV method ($M_{PV}$) using synthetic evolutionary models for intermediate to low-mass stars.\@ In \S 3, we test our method by applying it to a set of 14 objects of the Taurus-Auriga association (d$\sim$140 pc) with previous mass determinations by dynamical methods.\@  In addition, we will apply the method to multiple systems and determine the mass for each component.\@ 
The conclusions are given in \S 4. 

\section{PV method}
We will begin by calculating the NIR effect of disks and  show that the excess seen in CC and CM diagrams is similar to the behavior of the reddening vector produced by interstellar extinction. We will first illustrate this effect with a simple reprocessing disk via a simple and easy to manage calculation.\@ The real situation, however, is more complex.\@  Heating of the disk is also accomplished via viscous dissipation during accretion, and a plethora of parameters contribute to the process.\@  Thus, we present the NIR colors and magnitudes of accretion disk models \citep{dal05} in the following section.\@  Extinction is briefly described following that and then, we identify the principal vectors, describe how these vectors can be used to determine stellar masses and the expected limits of applicability of the method.

\subsection{Excess emision vector for reprocessing disks models}
Passive disks models do not have an intrinsic luminosity and only re-emit the energy absorbed from the central star.\@ Thus the superficial disk temperature can be approximated by a power law expression (like in \citealt{nat00} (hereafter N00)) like 

\begin{equation}
T(r_{D})\sim T_{0}\left(r_{D}/R_{*}\right)^{-q}, 
\end{equation}

where the $r_{D}$ is disk radius, $T_{*}$ and $R_{*}$ are the effective temperature and the stellar radius respectively.\@  N00 find that the parameters $T_{0}=0.75T_{*}$  and q=$0.47$ provide a good agreement with most observations. We considered a collection of concentric rings, each one emitting as a blackbody in agreement with eq 1.\@  The range of temperature goes from an inner radii with $T_{in}$=1500 K (dust sublimation temperature) to a maximum radii with $T_{out}$=100K. \@ Taking into account these considerations we calculate the predicted JHK disk flux. The total disk flux at a given frequency is given by, 

\begin{equation}
f[D]_{\nu}=\int_{\Omega} B_{\nu}(T(r_{D}))d\Omega,
\end{equation}  

\noindent where $B_{\nu}(T(r_{D}))$ is the Planck function and $T(r_{D})$ is the local temperature as a function of the disk radius.\@ 

So, to carry out the integral getting the next expression:

\begin{equation}
f[D]_{\nu}=\alpha\beta^{2}\kappa^{2/q} \int_{\gamma T_{in}}^{\gamma T_{out}}\frac{x^{-(2/q)+1}}{e^{1/x}-1}dx,
\end{equation} 

\noindent where $\alpha=4\pi h\nu^3/qc^{2}$, $\beta=R_{*}/d_{obs}$, and $\kappa=kT_{0}/h\nu$, the integral limits are given as a function of the disk temperature, being $T_{in}$ and $T_{out}$ the values previously mentioned and $\gamma=k/h\nu$.\@  To obtain results from eq. 3, we only need to know the effective temperature of the central star $T_{*}=T_{eff}$, the stellar radius $R_{*}$, and the frequency $\nu$. Once the disk flux is calculated, we add it to the stellar flux following the next expression that constitutes the star-disk system [s+d] 

\begin{equation}
f[s+d]_{\nu}= f[MS*]_{\nu}+\epsilon f[D]_{\nu}, 
\end{equation} 

\noindent where $\epsilon$ represents several parameters such as the inclination angle to the line of sight and circumstellar dust properties.

We use a set of intermediate to low mass main-sequence stars (Spectral Type from A0 to M6) and construct the s+d system using effective temperatures and bolometric magnitudes from \citet{sch82} to calculate the stellar radius and standard JHK photometric values from \citet{bes88}.\@ In Figures 1a and 1b we show the C-C (J-H vs. H-K) and C-M (K  vs. J-K) diagrams with the set of main sequence stars and the s+d systems.\@ The results are similar to those of others authors. \citet{lad92}, make use of C-C D to show how the IR colors are affected by the presence of circumstellar disks on intermediate to low mass stars, and achieved similar results.\@  Another useful diagram is the Color-Magnitude.\@ 
We use the K vs. J-K because with this combination we get the maximum contrast between the IR excess produced by the presence of the disk, which affects mostly K, and the interstellar extinction which has a greater effect in J.\@ 

\begin{figure}[!t]
\includegraphics[bb = 5 150 562 688, clip, width=\columnwidth]{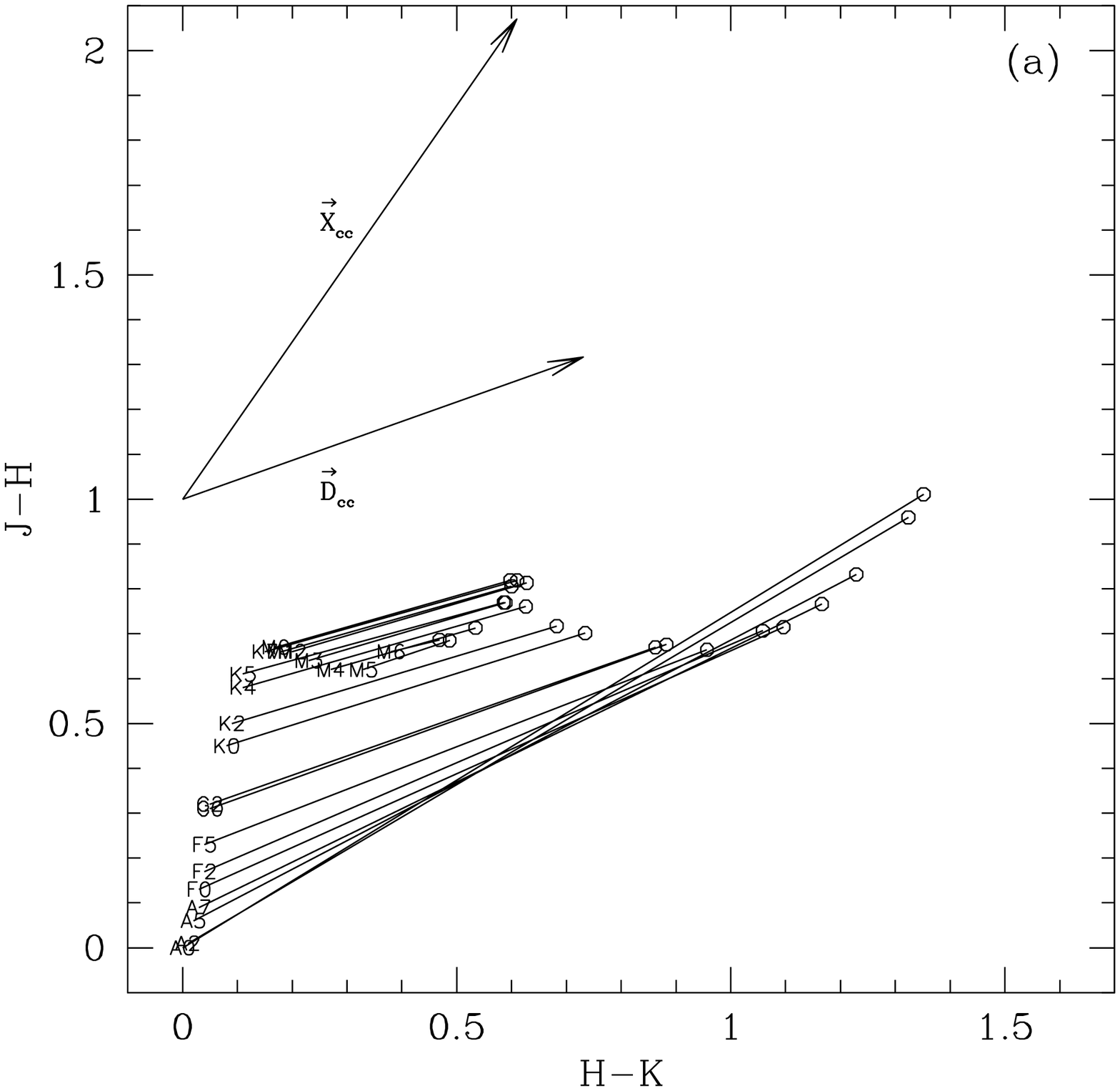}
\includegraphics[bb = 5 150 562 688, clip,width=\columnwidth]{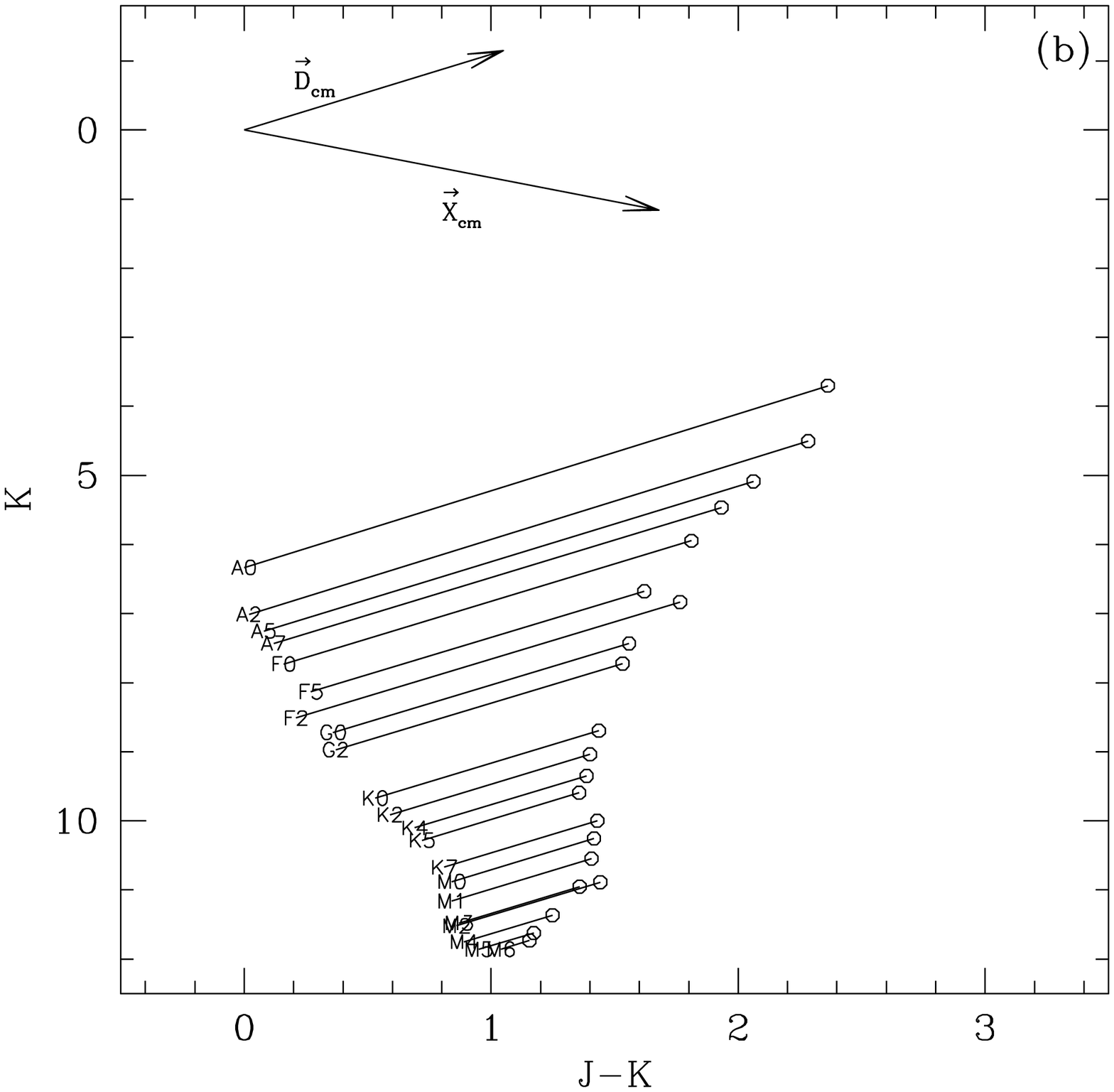}
\caption{(a)(J-H) vs. (H-K) and (b)(K) vs. (J-K) diagrams used to determine the PV generated by the disk reprocessing vector ($\vec{D}$) and interstellar extinction vector ($\vec{X}$).\@ We use stars with spectral types from A0 to M6, with a value of $\epsilon=0.4$ that corresponds to a $66\arcdeg$ of inclination from the line of sight .\@ The directions of the vectors are shown on the top left part on each figure.\@ }
\label{fig1}
\end{figure}

The main sequence is labeled with the spectral type of the star, the open circles represent the s+d model, the solid lines joining the two marks are the disk's vector direction and length, that is, the displacement produced by the presence of the disk.\@  
We show the case for $\epsilon=0.4$ that corresponds to a $66\arcdeg$ of inclination from the line of sight.\@   From figures 1a and 1b we perceive that it is possible to represent the IR excess produced  by the disk as vectorial quantity in each diagram.\@  This vector points in nearly the same direction for stars of all masses in the low to intermediate range.\@  We only start to see a discrepancy in the direction for stars earlier than A0.\@  
For the rest of the spectral types, the median value of the displacements, represented by $\vec{D}$, reproduces the mean effect of the disk.\@ The length of the vector is however different for each case, but this only implies that an amount $\Vert\vec{D}\Vert$ of disk emission should be interpreted differently for each spectral type.\@ This same procedure was repeated for a collection of 10,000 randomly oriented viewing angles from 0 to 90 $\deg$ .\@  The resultant median vector $\vec{D}$ shows little variation in the direction.\@ 
In The C-C D the slope of the excess emission vector is 0.43$\pm$0.107 and in the C-M D the slope is -1.097$\pm$0.02 .\@ Thus, the obtained components of the excess vectors are: $\vec{D}_{cc}$=(0.731, 0.317) and $\vec{D}_{cm}$=(1.05, -1.15) in magnitude units.\@

\subsection{IR excess vectors for accretion models}   
Following the modeling work of D'Alessio and collaborators (1998, 1999,  2001), they have published the SED's for a grid of 3000 accretion disks models that cover a wide range of parameters \citep{dal05}.\@  These models include accretion rates between $10^{-7}$ to $10^{-9}$ M$_\odot$/yr, onto central stars of 0.8 to 4 M$_\sun$ with temperatures from 4000 to 10000K and ages of 1 to 10 Myr.\@  Disk sizes range from 100 to 800 AU, and have inner radii that depend on the dust destruction temperature, taken around 1400 K, where the disk develops a wall whose scale height is also considered.\@  Gas and dust are well mixed in the disk, with dust size distributions given by power laws with indexes of 2.5 and 3.5 and maximum particle sizes of 1 $\mu$m to 10 cm.\@ The systems are viewed at two inclination angles; $\cos(i) = 0.5$ and $\cos(i) = 0.86$.\@

From this collection of models, we have chosen those that provide a clear view of the star ($\cos(i) = 0.86$) so that the effects of extinction can be separated from those of IR excess.\@ The effects of extinction will be addressed in the next section.\@  We also limit the following analysis to a particular age of 10 Myr for clarity.\@  Thus we are left with $\sim$ 730 disk models.\@  From the SED's of the star+disk systems we compute the magnitudes in the J, H and K bands (calibrating according to \citealt{bes88}) as well as the corresponding magnitudes for each central star, also given in \citet{dal05}.\@ 

\begin{figure}[!t]
\includegraphics[bb = 5 150 562 688, clip, width=\columnwidth]{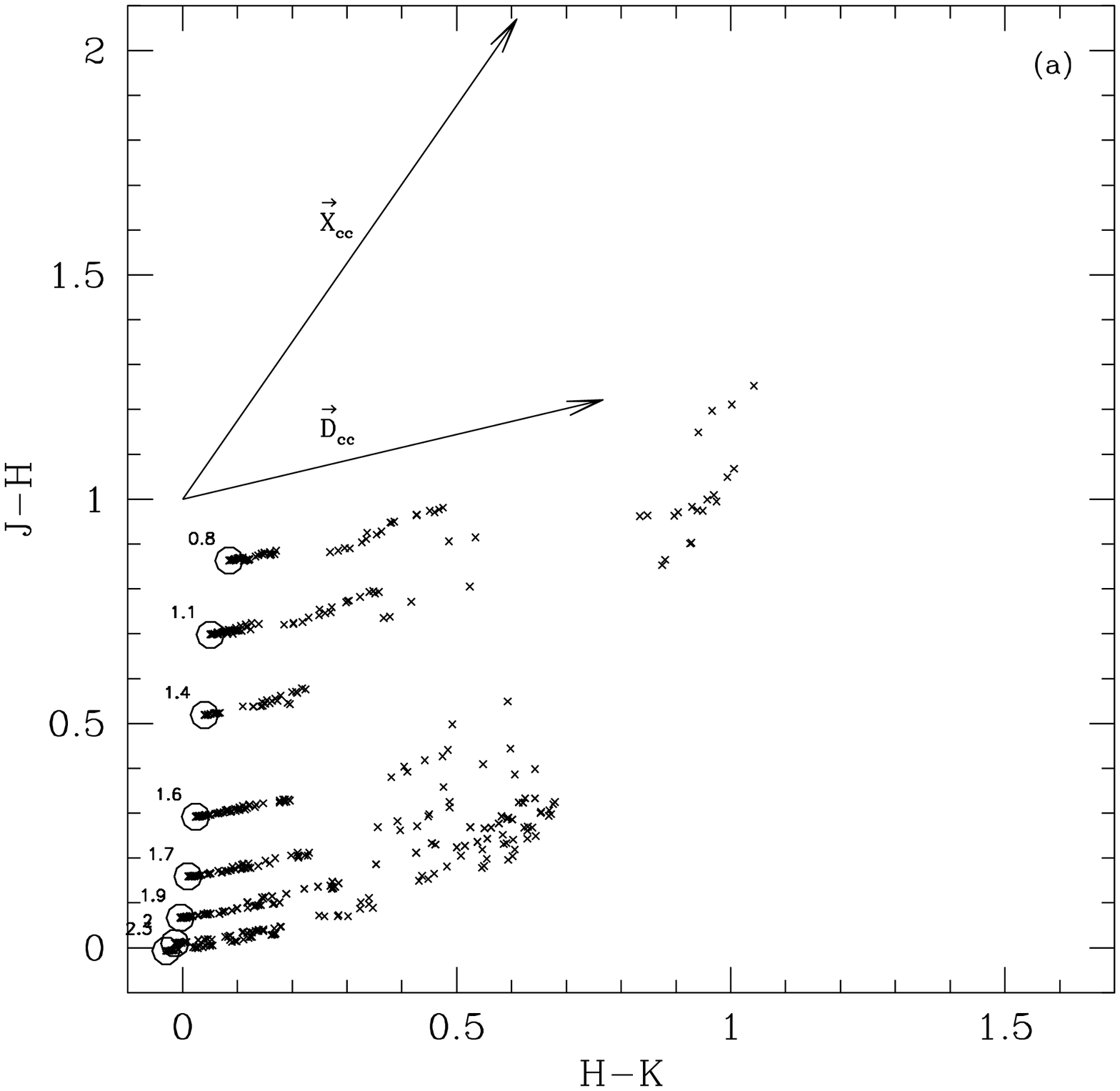}
\includegraphics[bb = 5 150 562 688, clip,width=\columnwidth]{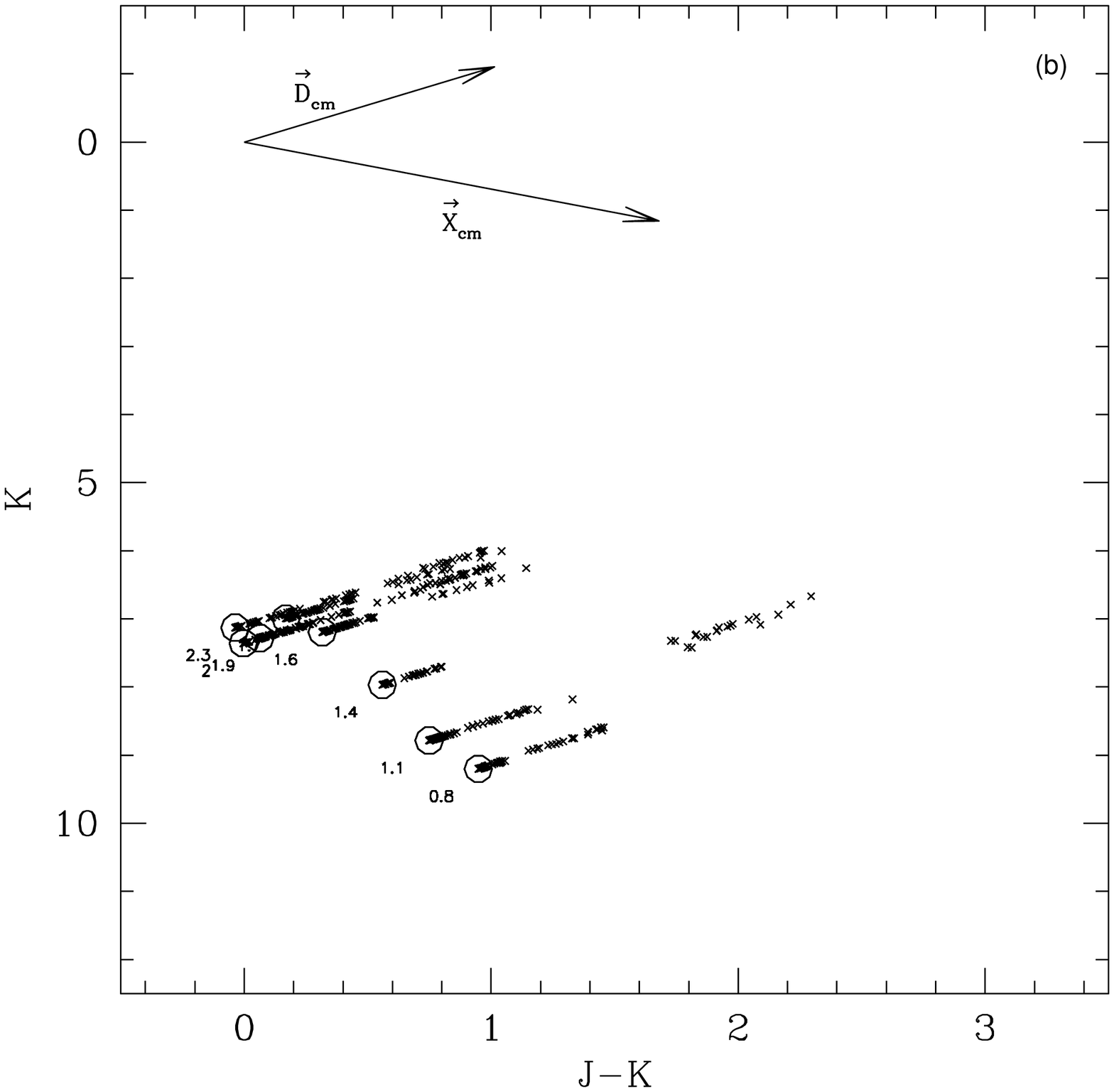}

\caption{(a)(J-H) Vs(H-K) and (b)(K) vs.(J-K) diagrams used to determine the PV.\@ The disk vector ($\vec{D}$) is generated by the median value of $\sim$ 730 accretion disk models and the interstellar extinction vector is represented by $\vec{X}$. }
\label{fig2}
\end{figure}

In Figure 2 we present the C-C and C-M diagrams.\@ In each panel the colors and magnitudes of the central star are represented by circles labeled with the stellar masses (0.8, 1.1, 1.4, 1.6, 1.7, 1.9, 2.0 and 2.3 $M_{\odot}$), and the crosses represent the corresponding  star$+$disk systems.\@  

The circles that corresponds to masses around 2 $M_{\odot}$ tend to pile up together, as the masses themselves are very close together.\@ However we have decided to keep the limits of the figure to allow a direct comparison with Figure 1.\@ In the C-M diagram it is evident that the disk models for each star occupy a very definite region, lying along a unique direction to the upper right, that is, going redder in J-K and brighter in K.\@  The median direction is indicated by an arrow, and it is very close to the direction found for simple reprocessing models (previous section).\@  In the C-C diagram, the disk models corresponding to each star also tend to occupy the upper right region, that is, redder in both colors, although the directions show a little more dispersion than in the previous case (reprocessing models).\@  In fact, the deviations from a common direction of all vectors in this figure is 0.15 radian, while only 0.02 radian for the C-M diagram.\@  Also note that the deviations are more important as the mass of the star increases, in the sense that the vectors curve upwards for shorter vectors.\@   This fact will limit the range of applicability of the method that we are proposing, since we will rely on a common direction for this excess vector, as indicated by the arrow in the C-C figure. Another important fact is that if we ratio the magnitude of each excess disk vector in the C-M diagram to it's corresponding magnitude in the C-C diagram, we get a constant 1.73 $\pm$ 0.14, indicating that vectors behave accordingly in both diagrams.\@  Thus, we may define the median disk vectors with components $\vec{D}_{cc}$=(0.767, 0.221) in the C-C D  and $\vec{D}_{cm}$= (1.014, -1.105) in the C-M D (actual median values have been arbitrarily multiplied by 13 to allow a better comparison), which are indeed very similar to the ones derived for the simple reprocessing disk models of the previous section.
 
\subsection{Extinction Reddening Vector}
Normal interstellar extinction can be represented as a reddening vector with components $[E(J-K),A_{K}]$ and $[E(H-K),E(J-H)]$  in the C-M and C-C  diagrams respectively.\@  
In Figures 1 and 2, we denoted by $\vec{X}$ the vector corresponding to $A_{V}=$10 and normal colors given in \citet{rie85}.\@ It is our intention to use these vectors to quantify the reddening produced by the ISM, the circumstellar envelope and the disk itself, on the photosphere of the star and the NIR emitting region of the disk.  However, we must proceed carefully in this task.\@  It is known that $R$, the ratio of total to selective extinction, attains larger than normal values in dense cores and dark clouds associated to star forming regions.\@ This is thought to arise because the shielded conditions of high opacity regions favor the coalescence of small grains into larger particles.\@  Larger particles in turn produce a more geometrical attenuation of visible and NIR light, resulting in a ``gray" extinction. 

In a recent paper, \citet{moo05} has observed the Hydrogen NIR recombination lines  towards UC-HII regions, to show that the 2$\mu$m opacity becomes flatter than normal only when $A_{V} > 25$ mag.\@  If an appropriate value of $R$ can be determined for these dense cores, the components of the reddening vector can be readily determined.\@ 

The mid planes of circumstellar disks are also regions of high opacity that can affect the observations of a central star with abnormal extinction.\@  However, \citet{dal01} argue that this is only a problem for a small ($15\%$) fraction of T Tauri stars in Taurus (a typical star forming region).\@ The increasing size in dust particles results in a growing critical viewing inclination at which the obscuring $A_{V} > 30$ condition is reached.

We can therefore assume that a normal extinction reddening vector is appropriate for $\approx$ $85\%$ of PMS stars in a SFR similar to Taurus.\@  Sources for which this is not the case (which should be a small number), can be pointed out by an unusually larger ($\sim$30 mag) extinction value.\@ For these sources, the extinction reddening vector would rotate in the C-M diagram to become nearly vertical, signaling that only attenuation of light and no reddening is present.\@ In the C-C diagram the vector would decrease in magnitude significantly, and even disappear, invalidating the application of the method that we are proposing.\@  This includes the most  extreme cases of a disk viewed edge on,  when the light of the star becomes completely extincted and the spectral energy distribution is dominated by scattered light from the polar regions of the disk.

\subsection{Principal Vectors}
From Figures 1 and 2 we perceive that it is possible to represent the IR excess produced  by the disk (either reprocessing or accretion) as vectorial quantity in each C-C and C-M diagram.\@ This vector points nearly in the same direction for stars of all masses in the low to intermediate range.\@ For masses grater that $\approx$ 2.5 M$_\odot$ or spectral types earlier than A0 the vector starts to rotate, so we regard this as the upper limit for the mass.\@  For lower stellar masses, the median value of the displacements, represented by $\vec{D}$,  reproduces the mean effect of the disk.\@ The magnitude of the effect is however different for each case, but this only implies that  an amount $\Vert\vec{D}\Vert$ of disk emission should be interpreted differently for each case, depending on the stellar mass.\@ To achieve a more realistic approach,  we will consider the $\vec{D}$  vectors corresponding to the accretion models that were derived in \S 2.2.\@ Furthermore, we note that the direction of $\vec{D}$ is very different from the reddening vector produced by extinction $\vec{X}$.\@  This vector seems appropriate for star forming regions not affected by dense cores of $A_V > 25$ mag  and disks not viewed at inclination angles such as $A_V > 30$ mag that would be affected by abnormal extinction.\@ 

Therefore we have a pair of vectors $\vec{D}$ and  $\vec{X}$ for each diagram.\@  We call them $\vec{D}_{cc}$ and $\vec{X}_{cc}$ for the C-C D and $\vec{D}_{cm}$ and $\vec{X}_{cm}$ for the C-M D.\@ The components of the vectors are: $\vec{D}_{cc}$=(0.767, 0.221), $\vec{X}_{cc}$=(0.6127, 1.064), $\vec{D}_{cm}$=(1.01, -1.105) and $\vec{X}_{cm}$=(1.676, 1.1613) in magnitude units. \@ These two sets of vectors are linearly independent so they represent an alternative basis for both diagrams.\@   By this way we can make new C-C and C-M diagrams using $\vec{D}$  and $\vec{X}$ as a  basis with respect to an arbitrary origin.\@

So if we have a diskless standard star standard without extinction (and at a given distance), it will occupy a certain position in each diagram.\@ If we now add certain amount for extinction and disk IR excess, it will move accordingly in each diagram.\@ 
On the other hand, if we start from a s+d system, we can subtract a certain amount of $\vec{D}$ and $\vec{X}$ to recover the original location of the star on the diagrams.
Moreover, if we do not know the original stellar mass (or spectral type), we can run the same procedure for all the possible standard stars (for example, the main sequence), and find in which case we get the same $\vec{D}$ and $\vec{X}$ values for the same standard star on each diagram.\@ In the next section, we will show that this procedure leads us to identify the star's mass with an acceptable error.\@ 

To proceed we need to locate the main sequence on the transformed diagrams.\@ And in the case that we have young stars we need to place the isochrones generated by  evolutionary models at a certain age.\@ This presents a problem since different models disagree up to factors of 2 in the masses predicted for the same age and spectral type.\@  We will consider several models (see \S 3.3) but in order to proceed here, we will start by using PS99 as a first example of the use of the method.\@ To plot the isochrones on the new diagrams first we need to know the absolute JHK magnitudes.\@  We use the results of \citet{tes98} for the bolometric corrections made to the PS99 evolutionary tracks.\@ Later on,  we transform the C-C and C-M diagrams coordinates to the new basis formed by the Principal Vectors (PV). 

\section{Application to a dynamically selected sample of YSO's}
To test the PV method, we use a sample of Taurus-Auriga (d$\sim$140 pc) PMS with stellar masses known by dynamical methods (TD02; \citealt{sch03}; \citealt{sim00}).\@ 
The majority of the Taurus-Auriga objects are intermediate to low mass and have a large fraction of binary or multiple systems.\@ According to the study made by \citet{ken95}, the objects are on an evolutionary status of no more than 2 Myr, showing a great amount of IR excesses (see \citealt{hai01}).\@  The star forming region is believed to be nearly coeval \citep{har01}, which is one of the advantages in studying stellar clusters, that we can assume that all the members have an age similar to the age of the cluster \citep{sca86}.\@ For this reason first we assumed an age of 2 Myr (log(t)=6.3 dex) for all objects. 

\begin{table}[ht!]\centering
  \setlength{\tabnotewidth}{\columnwidth}
  \tablecols{5}
  \caption{JHK photometry values for Taurus-Auriga objects that have mass values calculated by dynamical methods, in column 5 we identify if the object is multiple.}
  \begin{tabular}{lcccc}
    \toprule
    object & \multicolumn{1}{c}{J} & \multicolumn{1}{c}{H} & \multicolumn{1}{c}{K}&  \multicolumn{1}{c}{Multiple?}\\
    \midrule
BP Tau$^1$  &9.30 &8.42&	8.05&	No \\
CY Tau$^1$  &9.76 &8.90&	8.42&	No \\
DL Tau$^1$  &9.69 &8.77&	8.12&	Yes \\
DF Tau$^1$  &8.32 &7.40&	6.81&	Yes \\
DM Tau$^1$  &10.41&9.70& 9.45 &	No \\
FO Tau$^2$  &10.33&9.34&	8.81& Yes \\
FS Tau$^2$  &10.66&9.14&	7.74& Yes \\
GG Tau$^1$  &8.79 &7.85&	7.25&	Yes \\
GM Aur$^2$   &9.37 &8.73&	8.48& No \\
LkCa15$^2$   &9.51 &8.68&	8.22& No \\
UZ Tau E$^1$ &8.45 &7.60&	7.02&	Yes \\
V397 Aur$^3$ &9.21&8.46&	8.27& Yes \\
V773 Tau A$^1$&7.63 &6.83&	6.48&	Yes \\
ZZ Tau$^1$  &9.52 &8.78&	8.54&	No \\
    \bottomrule
 \multicolumn{5}{l}{\footnotesize 1:\citet{str89}}\\
\multicolumn{5}{l}{\footnotesize 2:\citet{ken95}}\\
\multicolumn{5}{l}{\footnotesize 3:\citet{wal88}}\\
\end{tabular}
\end{table}

Our dynamical sample consists of 14 objects (Table 1) that have JHK photometry reported by several authors (\citealt{str89}; \citealt{ken95}; \citealt{wal88}).\@ The extinction reported for these objects is generally small ($A_V<<30$), so it is not likely that any of them are viewed edge on, and so the PV method can be applied.\@ Some of the objects in the sample are known to be multiple systems.\@ For this first calculation we took the combined photometry of the systems even if they were multiple (we did not resolve the components, but see \S 3.5).\@ We applied the PV method to these objects using an isochrone of log(t)$=$6.3 dex.\@ We obtained two diagrams (Figures 3a and 3b) on the new D-X space generated by the PV.\@  The axis on both figures are in the same scale, so we can compare them directly.\@  All the distances on the D-X diagrams are measured in extinction units (eU) that corresponds to  $A_V=10$ or a disk unity.\@ We can see that if we take an object as a reference point on one graph and we put the other on top of it, there is a pair of mass values on the isochrones that superimpose or are very close together.\@ That is, we are finding the mass without disk or extinction, and from the vectors, the value of the direct contribution of the disk excess and interstellar extinction.\@ 

\begin{figure}[!t]
\includegraphics[bb = 5 150 562 688, clip,width=\columnwidth]{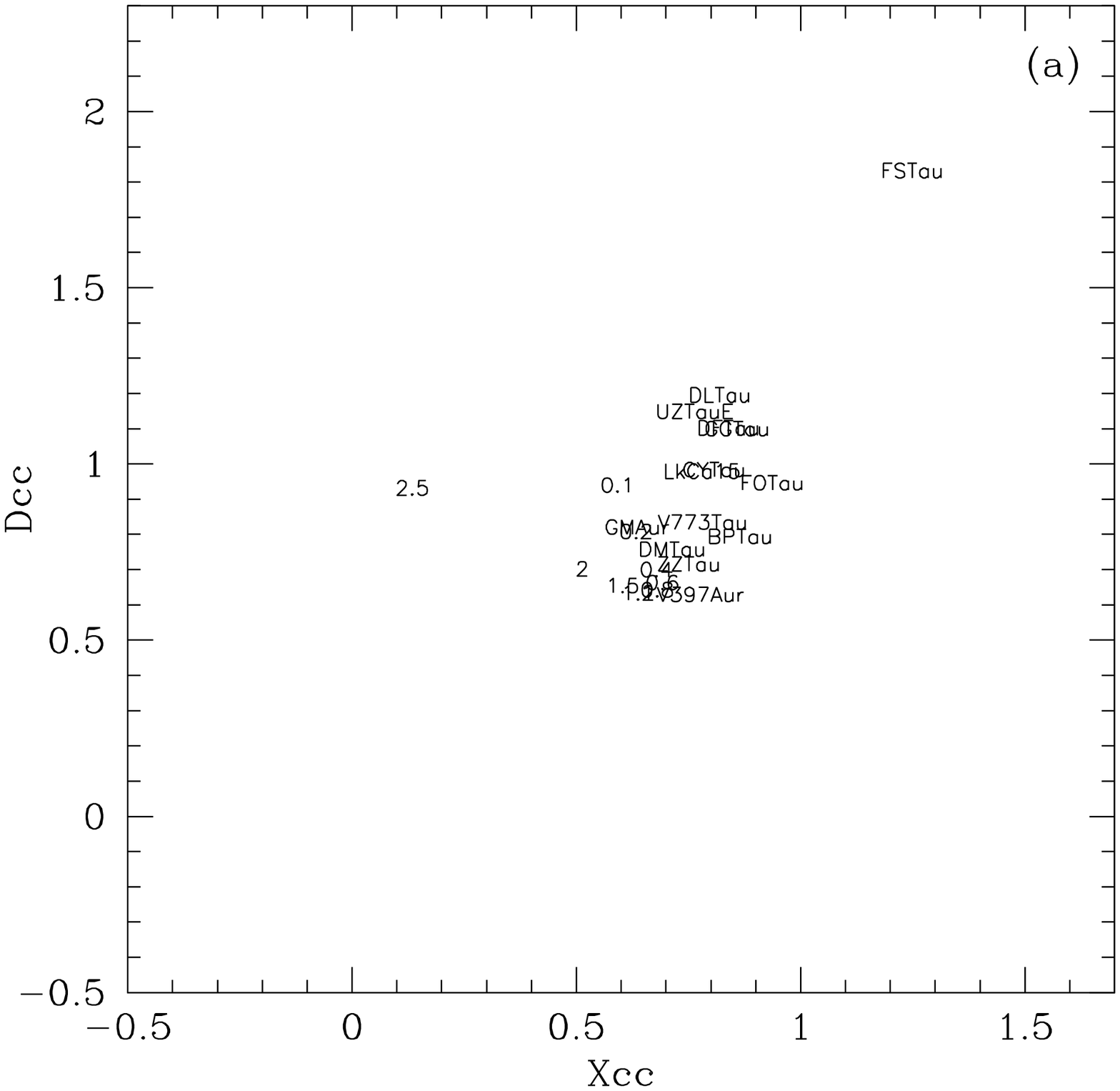}
\includegraphics[bb = 5 150 562 688, clip,width=\columnwidth]{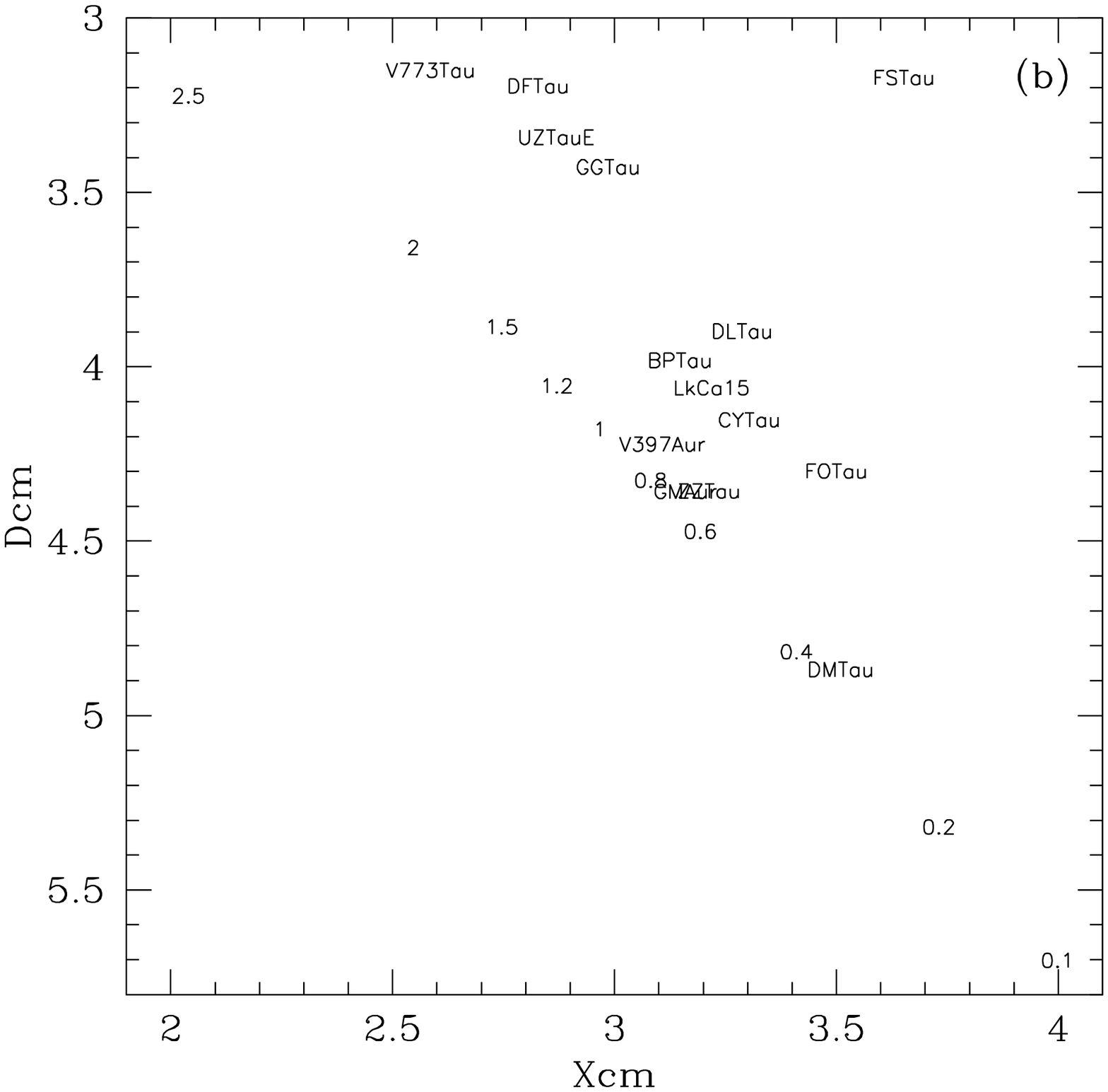}

\caption{D-X diagrams showing the location of the 14 Taurus-Auriga objects on the new space made by the PV.\@  The numbers represents the masses on a log(t)$=$6.3 dex isochrone.\@ The objects are identified by their names.\@ The axis units values are extinction units (eU).}
\end{figure}

\begin{figure}[!t]
\includegraphics[bb = 5 150 562 688, clip, width=\columnwidth]{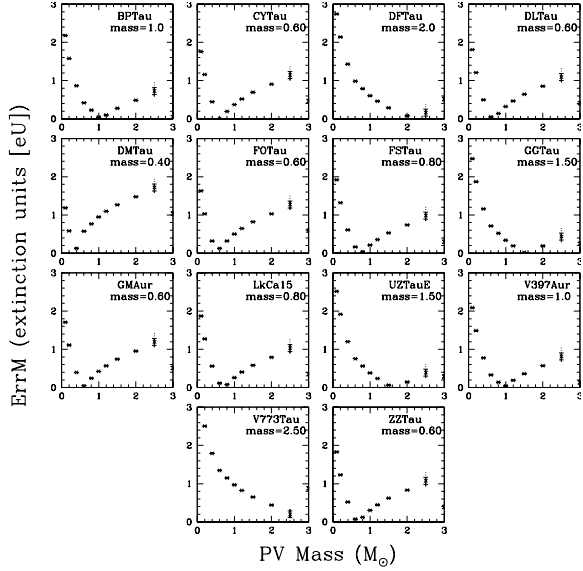}
\caption{$M_{PV}$ mass calculated for 14 objects at a given age of log(t)$=$6.3$\pm$0.05 from the Taurus-Auriga association.\@ Each graph is labeled by the name of the object and calculated mass given in $M_{\odot}$.\@ ErrM represents the minimum difference between the vectors length.\@  The units are extinction units (eU).}
\end{figure}

We point out that the pair of mass values on the two isochrones are separated by a distance, and this translates to an uncertainty value of the method.\@ So, to calculate the mass of the object, we select the mass pair with the "minimum difference" or "error measure" (ErrM) between them. \@

\begin{figure}[!t]
\includegraphics[bb = 5 150 562 688, clip, width=\columnwidth]{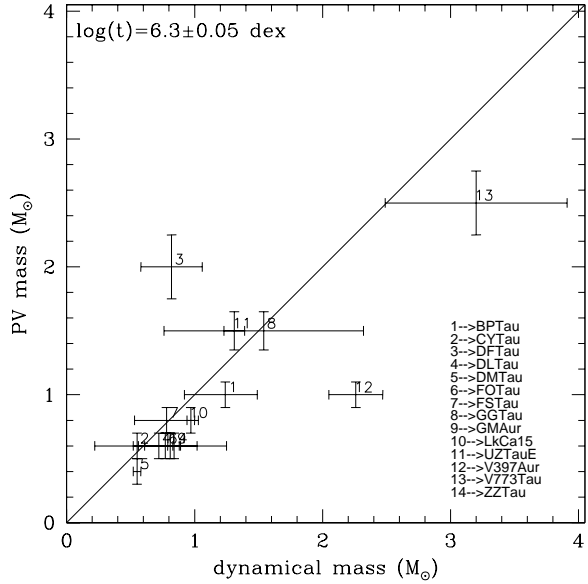}
\caption{$M_{PV}$ vs $M_{dyn}$ for the objects on a given age.}
\end{figure}

We can see this more clearly if we plot the ErrM vs. mass for each object.\@ In Figure 4 we can appreciate that the minimum ErrM value corresponds to the stellar mass.\@  The mass values are selected from a set of given values (0.1, 0.2, 0.4, 0.6, 0.8, 1, 1.2, 1.5, 2 and 2.5 $M_{\odot}$) that conform the reported isochrones by PS99.\@ We are not doing any interpolation to improve the mass precision.\@ This defines the uncertainty on our mass determination.\@  Furthermore, PS99 did not list temperature and luminosity values uniformly with respect to age for each considered mass.\@ Because of this, it was necessary to take an age dispersion of $log(t)=6.3\pm0.05$ dex, that corresponds to an uncertainty value in age of  $\sim0.3$ Myr.\@  In the cases where the same mass value appears several times on an any given age interval, we calculated the ErrM for each one and then obtained the mean value.\@ The variations with respect to the mean values are shown on Figure 4 as vertical bars on each point.\@  

\subsection{Comparing the PV method to dynamical mass values}
In the last section we obtained the mass values for 14 objects applying the PV method.\@ Now, we compare these PV masses with the values found by dynamical methods, using orbital information of binary systems (TD02; \citealt{sch03}) or circumstellar disk rotation kinematics \citep{sim00}.\@ 

 \begin{table}[!t]\centering
  \setlength{\tabnotewidth}{\columnwidth}
  \tablecols{8}
  % Stretch the space between table columns 
  \setlength{\tabcolsep}{0.5\tabcolsep}
\caption{Results obtained with the PV Method at a given age of log(t)$=$6.3$\pm$0.05 dex, for the Taurus-Auriga objects.\@  $M_{PV}$, disk vector magnitude value, interstellar extinction, and minimum ErrM value.}
\begin{tabular}{lcccccc}\toprule
Object&$M_{dyn}$&$\pm$err&$M_{PV}\pm$err&$\vec{D}$ &$\vec{X}$& ErrM\\	\midrule	
BP Tau  &1.24$^{2}$  &$_{-0.32}^{+0.25}$&1.0$\pm$0.1&0.192& 0.178&0.045\\
CY Tau  &0.55$^{2}$  & 0.33&0.60$\pm$0.1&0.112&0.320& 0.004\\
DF Tau  &0.82$^{4}$  & 0.24&2.0$\pm$0.25&0.303&0.431&0.078\\
DL Tau  &0.72$^{2}$  & 0.11&0.60$\pm$0.1&0.110&0.553&0.053\\
DM Tau  &0.55$^{2}$  & 0.03&0.40$\pm$0.1&0.067&0.004&0.127\\
GM Aur  &0.84$^{2}$  & 0.05&0.60$\pm$0.1&0.045&0.136&0.048\\
FO Tau  &0.77$^{4}$  & 0.25&0.60$\pm$0.1&0.275&0.227&0.126\\
FS Tau  &0.78$^{4}$  & 0.25&0.80$\pm$0.1&0.569&1.172&0.034\\
GG Tau  &1.54$^{0}$  & 0.78&1.5$\pm$0.15&0.247&0.452&0.019\\
LkCa15  &0.97$^{2}$  & 0.03&0.80$\pm$0.1&0.118&0.301&0.080\\
UZ Tau E &1.31 $^{2}$ & 0.08&1.5$\pm$0.15&0.141&0.519&0.062\\
V397 Aur&2.26$^{3}$  & 0.21&1.0$\pm$0.1&0.129&0.023&0.044\\
V773 Tau&3.20$^{4}$  & 0.71&2.5$\pm$0.25&0.596&0.011&0.200\\
ZZ Tau  &0.81$^{1}$  &$_{-0.25}^{+0.44}$&0.60$\pm$0.1&0.038&0.084&0.074\\\bottomrule
\multicolumn{7}{l}{\footnotesize 0:\citet{gui99};\citet{whi99}}\\
\multicolumn{7}{l}{\footnotesize 1:\citet{sch03}}\\
\multicolumn{7}{l}{\footnotesize 2:\citet{sim00}}\\
\multicolumn{7}{l}{\footnotesize 3:\citet{ste01}}\\
\multicolumn{7}{l}{\footnotesize 4:\citet{tam02}}\\
\end{tabular}
\end{table}

Table 2 (see also figure 5) shows this comparison.\@ An overall difference of 29$\%$ with dynamical masses can be appreciated.\@ However, for most cases the mass value fits the dynamical value within the dynamical error margins.\@ Of the 14 objects,  the uncertainty is less than 1$\sigma$ in 6 cases, less than 1.5$\sigma$ in 4 more and less than 3$\sigma$  in 2 more.\@  The last 2 cases are V397 Aur ($\Delta M=1.26$) and DF Tau ($\Delta M=1.18$). \@ We will explore possible causes for the discrepancies.\@

\begin{figure}[!t]
\includegraphics[bb = 5 150 562 688, clip,width=\columnwidth]{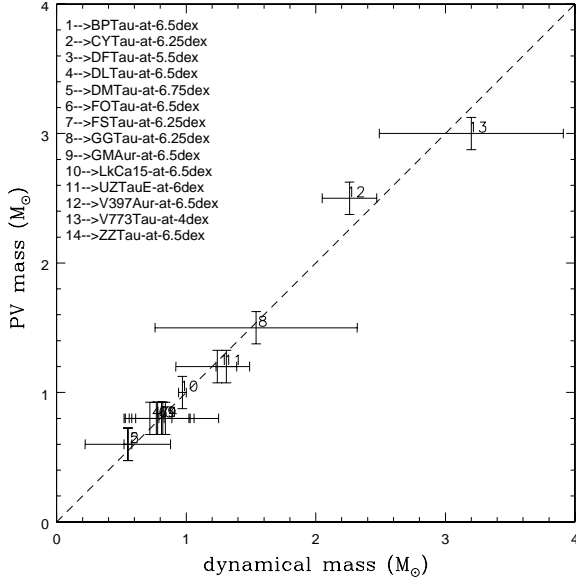}
\caption{$M_{PV}$ vs $M_{dyn}$ for the objects at the age of better match (listed).}
\end{figure}
 
There is a controversy on distance on DF Tau.\@ Hipparcos measurements \citep{fav98} place it at a distance between 31-52 pc.\@ \citet{lam01} based on HST measurements debate this value and place it at 70 pc.\@ However, \citet{sch03}, determine the system's dynamical mass to be 0.9 $M_{\odot}$ assuming a distance of 140 pc, the same distance used by TD02.\@ On the other hand, if the distance were as small as 70 pc, DF Tau would not be a Taurus member (extension of $\sim$20 pc), and the dynamical mass would change as $d^3$.\@ So we will take the fiducial distance of 140 pc for DF Tau.\@ 
V397 Aur (NTT 045251+3016) is a spectroscopic binary \citep{wal88} and is classified as a naked T Tauri.\@ The distance is also a problem, \citet{ste01} calculated their dynamical masses setting the system at 145$\pm$8 pc. \@ If we took DF Tau out of the comparison we would obtain a 20$\%$ agreement with dynamical masses.\@ 

We remark the fact that the calculation has been made taking a given age (coeval).\@ However there are indications that star clusters have an age dispersion.\@ For the Taurus region several authors have mentioned that this variation is $\pm0.3$ dex on age (\citealt{har03}; \citealt{pal02}).\@ This can be another issue to take into account to try to explain the observed differences.\@ To prove this last hypothesis, we will apply the PV Method backwards, that is, since for these objects we know the dynamical mass value,  we will find the  isochrone that gives the minimum ErrM for that mass.\@ In the next section we will address the question of obtaining mass and age simultaneously.\@ The outcome age values are listed in Table 3.(See also Figure 6).\@ There we compare them to the ages estimated  by other authors.\@ We appreciate in general a certain dispersion in the ages determined by other authors and in particular for the age of  DF Tau, all seem to agree that the age is significantly younger than log(t)$=$6.3 dex.\@ We get a  value of log(t)$=$5.55 dex, which is similar to the one previously determined by \citet{pal02} that assigned it an age of log(t)$=$5 dex, so probably DF Tau is younger than the cluster's age.\@  
For V397 Aur the age value found was log(t)=6.5 dex that corresponds to the upper limit for which we have data values on the evolutionary track for masses more than 2.5 $M_{\odot}$ (see also \S 3.2), so this object is out of the boundaries of the restrictions imposed by the method (more massive and older).\@   

\begin{figure}[!t]
\includegraphics[bb = 5 150 562 688, clip,width=\columnwidth]{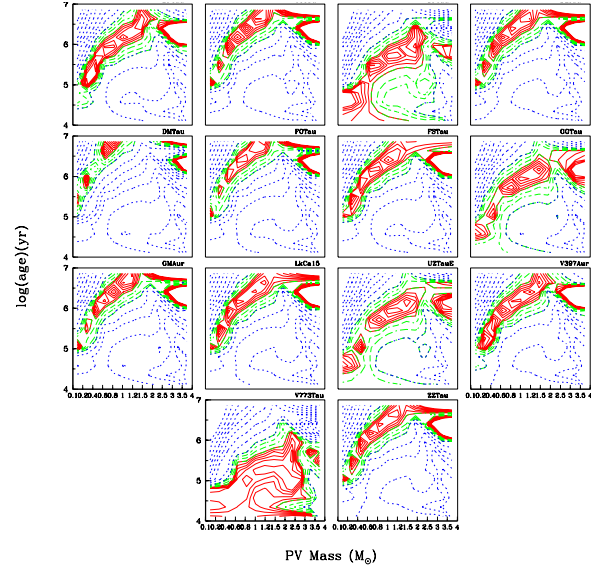}
\caption{ErrM as a function of mass and age for each object in the sample.\@ The solid line represents the region where ErrM has a minimum ($<$0.3 eU with a step of 0.05) value and this constitute possible solutions for mass and age.\@ The dashed line corresponds to 0.3 eU$<$ErrM$<$0.6 eU with a step of 0.1 and the dotted lines to an ErrM value greater than 0.6 eU (step of 0.25).}
\label{fig7}
\end{figure}

\begin{table*}[!t]\centering
  \setlength{\tabnotewidth}{2.9\textwidth}
  \setlength{\tabcolsep}{1.00\tabcolsep}
  \tablecols{11}
  \caption{ Ages determined by different authors, compared with PV obtained values.} \label{tabl3}  
  \begin{tabular}{lcccccccccc}
    \toprule
Object&$M_{dyn}$&$M_{PV}$&PS02&WG01&S00&St01&H95&T01&G04&$Age_{PV}\pm$0.12\\\midrule
BP Tau  &  1.24   & 1.0   &  6.25  &6.16&6.3-7.0&\nodata&	5.78 &\nodata& \nodata& 6.5 \\
CY Tau  &  0.55   & 0.60 &  6.25  &6.48&6.3-6.7&\nodata&	6.27 &\nodata&\nodata & 6.25\\
DF Tau  &  0.82   & 1.0   &  5.0   & \nodata    &\nodata&\nodata&3.74   &\nodata&\nodata&5.5\\
DL Tau  &  0.72   & 0.80  &  6.4   & $<$6    	 & 6.30  &\nodata&5.78   &\nodata&\nodata& 6.5 \\
DM Tau  &  0.55  &0.60  &  6.44  &6.80        & 6.70	 &\nodata&\nodata&6.4	&	 6.4	 &6.75\\
FO Tau  &  0.77   &0.80   &  5.6   &\nodata     &\nodata&\nodata&\nodata&\nodata&\nodata&6.5\\
FS Tau  &  0.78   &0.80   &  6.44  & \nodata    &\nodata&\nodata&\nodata&\nodata&\nodata&6.25\\
GG Tau  &  1.54  &1.5   &6.1-6.36&	\nodata	   &6.0-6.3&\nodata& 4.82  &\nodata&5.8-6.2 &6.25\\
GM Aur  &  0.84   & 0.80 &  6.98  &    6.88    & 6.47  &\nodata&	5.06 &6.25	&6.1-6.25&6.5 \\
LkCa 15 &  0.97   &0.80   &  6.71  &    6.80    &6.4-6.7&\nodata&\nodata&7.07 &6.9-7.07&6.75 \\
UZ Tau E &  1.31  &1.5   &  5.3   &  \nodata   &\nodata&\nodata& 3.60  &\nodata&\nodata&6.0\\
V397 Aur&  2.26  &2.5   &  6.52  &  \nodata   &\nodata&6.78   &\nodata&\nodata&\nodata&6.5\\
V773 Tau&  3.20 &3.0  &6.44-6.6&  \nodata   &\nodata&\nodata&\nodata&\nodata&\nodata&4.0 \\ 
ZZ Tau  &  0.81  &0.80  &  5.6   &  \nodata   &\nodata&\nodata&\nodata&\nodata&\nodata&6.5 \\ \bottomrule
\multicolumn{11}{c}{PS02:\citet{pal02}; G04:\citet{gre04}; H95:\citet{har95}}\\ 
\multicolumn{11}{c}{WG01:\citet{whi01}; S00:\citet{sim00}; St01:\citet{ste01}}\\ 

\end{tabular}
\end{table*}

\subsection{Mass and Age}
We have seen that we can obtain different mass values with the PV method if we change the age.\@ We want to explore if it is possible to obtain the age and mass simultaneously for each object.\@ That is to develop a method to determine the two quantities.\@ 
However, this solution involves 4 quantities (mass, age, size of the disk and extinction reddening) and we only have 3 photometric measurements (J, H and K).\@ So it is expected that the solution is not unique, and dependent of one of the parameters.\@
We apply the PV method for a grid made by age values and visualize the results as a contour map with the ErrM error as a function of two variables (mass and age).\@ We can see how the minimum value changes as a function of mass and age.\@ These maps are shown in Figure 7.\@  We can detect the presence of a region on the top left of each panel (continuous solid line) where a canyon like shape is formed.\@ This region is generated by all the minimum ErrM that corresponds to all the possible acceptable solutions for mass and age for the object that are consistent with the measurements of J, H and K.\@

For DF Tau and V773 Tau (Figure 7) we can better appreciate that the contour figures are slightly different  from the rest of the objects.\@ The solid line zone is located at a lower position, indicating that these objects are younger than the rest, with ages less than $10^{6}$ yr for any given object's mass.\@  On the other hand, DM Tau shows the solid line zone located on a higher zone indicating an older age.\@  For V397 Aur, the zone extends to an older region as the mass is incremented, getting out of the range when the mass is larger than $\sim$2$M_{\odot}$, restricting us from finding the real mass  as we already mentioned.\@

\subsection{Choosing  a different evolutionary models}
We are bound to evolutionary track models, because those are used as a reference points by the PV method in the D-X space.\@  We explore the effect on mass determination if we take different evolutionary track models. In order to address this point, we used several evolutionary models such as  DM97, BCAH98, S00 and $Y^{2}$01, each one with bolometric corrections from different authors (\citealt{ken95} ; \citealt{lej98}; \citealt{sie97}) and we calculate the resulting $M_{PV}$ masses using the coeval age of log(t)=6.3 dex.\@  Then we compare the results for each model with dynamical masses in table 4. \@  In all the cases we took DF Tau out of the comparison for the reasons explained above.\@
For the mass values obtained with DM97 evolutionary models for a  mass range from 0.017 to 2.5 $M_{\odot}$  we obtained a difference of  23$\%$.\@   For the BCAH98 tracks we did not take into account  V773 Tau and V397 Aur because they have dynamical masses  greater than the superior mass limit  that is possible to determine with them (0.07 to 1.4 $M_{\odot}$), obtaining a value of 33$\%$ of variation with respect to dynamical masses.\@ 

The S00 evolutionary models consider an accretion of  $M_{acc}$ = $1.0x10^{-7}$ $M_{\odot}/yr$ over a period of $5x10^{6}$ yr onto the central star that produces an extra luminosity and an increase of mass.\@ For masses smaller than 0.4 $M_{\odot}$ a problem exists because the  accretion processes  directly affects the object, causing that an object with an initially mass of 0.2$M_{\odot}$ will appear on the evolutionary track in a locus of a 0.5$M_{\odot}$ star of the corresponding age.\@  After applying the method we obtained a difference of 34$\%$ for the model with overshooting and  45$\%$ without overshooting.  If we use the $Y^{2}$01 for a Z=0.02 and mass range from 0.40 to 2.6 $M_{\odot}$ we obtained a variation of 23$\%$ with respect to dynamical.\@  

We thus see that the evolutionary models proposed by DM97, $Y^2$, and PS99 give all similar values that compare to dynamical masses at the $\sim20\%$ level.\@ Furthermore, DM97 and $Y^2$ compare also reasonably good ($\sim23\%$ ) to PS99 tracks that we employed through \S 3.

Therefore, at this level of certainty we consider any of these models as equivalent, and we will continue to use the PS99 tracks.\@ However, we must emphasize that the PV method is very sensitive to the particular evolutionary model employed, and since this topic is not yet settled, people should remain aware of further developments in this area. 

\subsection {Uncertainties}
Several other factors have to be considered to put out the method into context, such as  the effect of distance, age uncertainties.\@ In this section we estimate these and other factors and quantify how much the PV mass value for an object is modified in each one of the listed conditions. We first explore variations in the distance to the object.\@  That is to say, if we assume that the region is not at 140 pc.\@  In order to see this, we made an analysis varying the distance in $\pm$ 10 pc ($7\%$), and  we applied the PV method with the new parameter to the sample objects.\@  The masses obtained yield a median difference of 22$\%$ with respect  to the original PV value at 140 pc. To examine the uncertainty that appears when we assign a different  age, we took PS99 tracks and we varied the age in $\pm$ 0.1 dex (22$\%$) and obtained a 18$\%$ of variation in the $M_{PV}$ determination with respect to the coeval age of  log(t)$=$6.3 dex. 

 \begin{table}[!t]\centering
\setlength{\tabnotewidth}{\columnwidth}
  \tablecols{8}
  % Stretch the space between table columns 
  \setlength{\tabcolsep}{0.4\tabcolsep}
\caption{Mass obtained using several evolutionary models. Age log(t)$=$6.3$\pm$0.05 dex. The units are in solar masses. $\sigma_{dyn}$ is the mean relative difference to dynamical mass. } \label{tbl2}
\begin{tabular}{lccccccc}\toprule
Object&$M_{dyn}$&	$M_{PS99}$&	$M_{s00}^{1}$&$M_{s00}^{2}$&$M_{DM97}$&$M_{BCAH98}$&$M_{YY}$\\\midrule	
BP Tau	&1.24	&1.0&1.0&	1.0&	1.30&0.80&1.10\\
CY Tau	&0.55&0.6&	0.60&	0.6&	0.80&0.45&0.60\\
DF Tau	&0.82$^{*}$&2.0$^{*}$&	1.89$^{*}$&3.5$^{*}$&	2.5$^{*}$	&1.1$^{*}$& 5.0$^{*}$\\
DL Tau	&0.72	&0.6	&0.60&0.6&	0.80&0.5&0.60\\
DM Tau&0.55	&0.4	&0.13&0.1&	0.40&0.30&0.5\\
FO Tau	&0.77	&0.6	&0.16&0.5&	0.60&0.40&0.5\\
FS Tau	&0.78	&0.8	&0.25&0.2&	1.0&0.57&0.80\\
GG Tau	&1.54	&1.5	&1.29&1.3&	2.0&1.15&1.80\\
GM Aur	&0.84	&0.6	&0.5	&0.5	&0.80&0.45&0.40\\
LkCa 15&	0.97&	0.8&	0.25&	0.6&	0.80&0.5&0.80\\
UZ Tau E	&1.31&1.5&	1.39&1.4&	2.0&1.40&1.90\\
V397 Aur&2.26&1.0&0.80&0.8&	1.2&0.60$^{**}$&1.0\\
V773 Tau&	3.2&	2.5&	6.0&	2.5&	3.0&1.10$^{**}$&	2.90\\
ZZ Tau&	0.81&	0.6&	0.60&0.6&	0.80&	0.5&	0.60\\\bottomrule
$\sigma_{dyn}$ ($\%$)& &20.29& 44.9 &34.6 &22.94 &33.64 &23.6\\
\multicolumn{8}{l}{\footnotesize 1:without overshooting}\\
\multicolumn{8}{l}{\footnotesize 2:overshooting}\\
\multicolumn{8}{l}{\footnotesize *:Not taken into account in the $\sigma_{dyn}$ value,  see text.}\\ 
\multicolumn{8}{l}{\footnotesize **:Out of mass range defined by the BCAH98 model (0.07 to 1.4 $M_{\odot}$). }\\
\end{tabular}
\end{table}

Another source of uncertainties could be due to the principal vectors themselves.  If a different value of R is assumed, the slope of $\vec{X}$ vector will only change in the C-M D.\@  But it will remain the same in the C-C D.\@  To quantify the magnitude of the PV mass produced by this change of slope we tried values of R (3.2 and 4) instead of the normal value of R$=$3.1 for the ISM.\@ The resulting PV masses differ in 1.5$\%$ and 37$\%$ respectively.\@ Therefore,  small change in the value of R has no significant effect on mass but a larger change such as required for dense cores (R$\sim$4) should be carefully considered.\@ 

Next, we want to evaluate the change in PV mass when the slope of the excess vector $\vec{D}$ varies.\@  We use the slope value of 0.58 obtained by \citet{mey97}  for the locus of the T Tauris as the slope value for the excess vector $\vec{D}$.\@ We assigned the magnitude $\Vert\vec{D}\Vert$ obtained through the models of  \citet{dal05}  of $\Vert\vec{Dcc}\Vert=0.798$. With these, values we applied the coordinate transformation to the D-X space at a given age of log(t)$=$6.3 dex.  We needed two diagrams to find the mass values  but \citet{mey97} obtained his T Tauris locus only in C-C D.  So, we decided to use the C-M D slope for the excess vector that was obtained with the \citet{dal05} accretion models, which has a slope of -1.09.\@  When we changed these parameters, we obtained a mass variation of 12$\%$ with respect to the values obtained in \S 3.1. 

\subsection{Binaries}
Many of the objects in our list belong to multiple systems and we have analyzed them as a single object.\@ So the next question emerges.\@ Can we apply the method to the components of a system and find the individual masses in a consistent way? To answer this, we need to know individual photometry for the components and apply the PV method for each one an determine the mass for a given age.\@ However, we have dynamical mass values only for V773 Tau, the other systems have their masses calculated by evolutionary tracks (\citealt{whi01})(hereafter WG01), and are taken so that the sum of the components match the mass values obtained by dynamical methods.\@ We decided to apply the PV method to the systems components and compare these values with mass values determined by evolutionary tracks.\@ Only GG Tau, FO Tau, FS Tau, and V773 Tau, meet the requirements as they have complete component photometry and determined masses (see table 5).\@ For the preliminary determination we fixed the age to log(t)$=$6.3 dex for all the systems, following the work of \citet{har94} where they studied 39 binary systems in Taurus-Auriga and Orion region, finding that 2/3 of the sample were coeval.\@ GG Tau is a quadruple interactive system \citep{lei91} of two binary pairs identified as GG Tau A and GG Tau B separated by a distance of 10."1 (1,414 AU).\@ For the GG Tau A components the separation is 0."25 (35 AU)( named GG Tau Aa and GG Tau Ab), (see mass values in table 6).\@ The GG Tau B components are separated by 1."48 (207 AU).\@  In Table 4 we show the JHK photometric values reported by \citet{whi99} and  \citet{ghe97}.\@  

The FS Tau system (Haro 6-5 A) has two objects separated by a distance of 0."25, and the secondary component is 12 times weaker than the primary \citep{che90}.\@ TD02 determined for the first time the dynamical value for the system's mass.\@  

V773 Tau is a multiple system, at distance of 148 pc \citep{les99}.\@ \citet{wel95} determined it as triple system with a very close pair (0.3AU) and the other component separated at 0."15 ($\sim$22 AU).\@  But in 2003, several authors reported a fourth component (\citealt{duc03}; \citealt{woi03}).\@  \citet{duc03} renamed the closer binary pair (System V773 Tau AB) as V773 Tau A and V773 Tau B and also V773 Tau C for the one located at 0."15 and V773 Tau D for the new component that was cataloged as an IR Source.\@

 \begin{table}[!t]\centering
  \setlength{\tabnotewidth}{\columnwidth}
  \tablecols{3}
  % Stretch the space between table columns 
  \setlength{\tabcolsep}{0.3\tabcolsep}
  \caption{Multiple systems Photometry.} 
  \begin{tabular}{lccccccc}
    \toprule
Components  &  J 	&$\Delta$J&  H  &$\Delta$H&  K   &$\Delta$K\\\midrule
GG Tau Aa$^{1}$ & 9.24& 0.09    & 8.27& 0.09    & 7.73 &  0.05\\ 
GGTau Ab$^{1}$	&10.12& 0.02    & 9.07& 0.08    & 8.53 &  0.08\\ 
GG Tau Ba$^{1}$	&11.48& 0.16    &10.63& 0.15    &10.20 &  0.12\\ 
GG Tau Bb$^{1}$ &13.16& 0.12    &12.38& 0.06    &12.01 &  0.22\\ 
FS Tau A$^{2}$  &10.85& 0.0     & 9.32& 0.0     & 7.78 &  0.0\\
FS Tau B$^{2}$  &12.66& 0.0     &11.17& 0.0     &10.03 &  0.0\\
FO Tau A$^{2}$  &10.13& 0.0     & 9.35& 0.0     & 8.76 &  0.0\\
FO Tau B$^{2}$  &10.95& 0.0	    & 9.59& 0.0     & 9.14 &  0.0\\
V773 Tau AB$^{2}$&7.77 & 0.05   &7.03 &0.03     &6.77 &0.09\\ 
V773 Tau C$^{2}$&10.13& 0.09    &8.91 &0.15     &8.09 &0.31\\\bottomrule 
\multicolumn{7}{l}{1:\citet{whi99}; 2:\citet{woi01}}
\end{tabular}
\end{table}

\begin{figure}[!t]
\includegraphics[bb = 5 290 562 688, clip,width=\columnwidth]{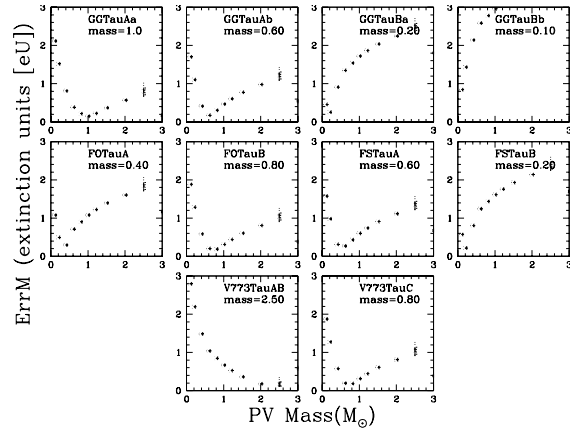}
\caption{Determined PV mass for the components of GG Tau, FS Tau , FO Tau and V773 Tau.\@ The PV Mass values are shown on the top right margin under the objects name on each graph.\label{fig8} }
\end{figure}

In Table 6 and Figure 8 we show the resulting mass values with the PV Method at a given age of log(t)$=$6.3 dex.\@ In general we can see that the method converges for most of the sample objects.\@  Only for the case of GG Tau Bb we can not find the minimum because we are mass limited.\@ We predict for this component a mass of $<$0.1 $M_{\odot}$ having an ErrM of 0.84 eU.\@ For the 10 of the objects listed in table 6 (Figure 9), 7 coincide better than 1.5$\sigma$ according to track mass uncertainties  and 1 more is closer than 2$\sigma$.\@ Only 2 cases are $>$3$\sigma$ away values (GG Tau Aa and FO Tau B).\@  We ran the PV method with a varying age to try to find if we can explain these discrepancies by finding the objects age.\@  We show the results in table 7 and Figure 10.\@  We can see that we can obtain a fitting value within 1$\sigma$, if the ages of GG Tau Aa and FO Tau B are log(t)$=$6.0 dex.\@  The age determined for GG Tau Aa matches the estimated by WG01 and for  FO Tau B we are within the uncertainties.\@ 

\begin{figure}[!t]
\includegraphics[bb = 5 150 562 688, clip,width=\columnwidth]{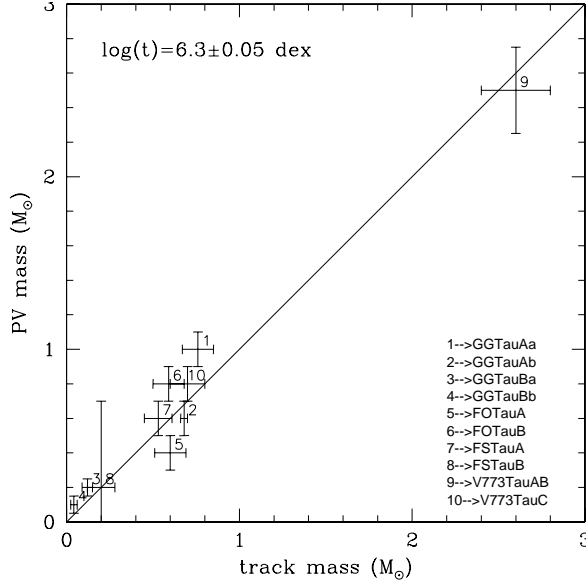}
\caption{ $M_{PV}$ vs $M_{Track}$ for the system components of GG Tau, FO Tau and FS Tau. The masses for the system V773 Tau are dynamical determinations made by \citet{duc03}. We assume a coeval age of log(t)$=$6.3$\pm$0.05 dex for all the systems.\label{fig9}}
\end{figure} 

\begin{table}[!t]\centering
  \setlength{\tabnotewidth}{\columnwidth}
  \tablecols{6}
  % Stretch the space between table columns 
  \setlength{\tabcolsep}{0.3\tabcolsep}
\caption{Mass determination for multiple systems compared to given mass by evolutionary tracks (WG01). PV mass values shown on column 3. Columns 4, 5 and 6 display the calculated disk and  interstellar extinction vectors and ErrM.}
\begin{tabular}{lccccc}\toprule
Object&$M_{Track}$$\pm$err&$M_{PV}\pm$err&$\vec{D}$&$\vec{X}$& ErrM\\\midrule
GG TauAa& 0.76$\pm$0.09 &	1.0 $\pm$0.1 &0.194&0.294&0.150\\
GG TauAb& 0.68$\pm$0.02	&	0.60$\pm$0.1&0.257&0.172&0.174\\
GG TauBa& 0.12$\pm$0.03	&	0.20$\pm$0.05&0.148&0.0191&0.255\\
GG TauBb& 0.042$\pm$0.019& 0.10$\pm$0.05&0.285&0.462&0.844\\
FO TauA & 0.60$\pm$0.09 & 0.40$\pm$0.1 &0.004&0.378&0.296\\
FO TauB & 0.59$\pm$0.09 & 0.80$\pm$0.1&0.648&0.236&0.188\\
FS TauA & 0.53$\pm$0.08 & 0.60$\pm$0.1&0.497&1.240&0.270\\
FS TauB & 0.20$\pm$0.08 &	0.20$\pm$0.5 &0.599&0.556&0.217\\
V773 TauAB&2.6$^{1}$$\pm$0.2&2.50$\pm$0.25 &0.590&0.266&0.175\\
V773 TauC &0.7$^{1}$$\pm$0.1&0.80$\pm$0.1&0.374&0.466&0.187\\\bottomrule
\multicolumn{6}{l}{1:dynamical mass by \citet{duc03}}
\end{tabular}
\end{table}

\begin{figure}[!t]
\includegraphics[bb = 5 110 562 688, clip,width=\columnwidth]{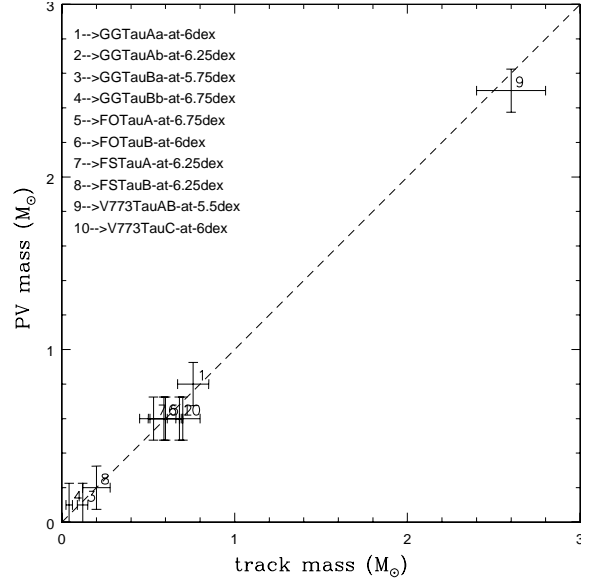}
\caption{ $M_{PV}$ vs $M_{track}$ to determine ages for the systems components at given mass that fits the mass value determined by evolutionary tracks.\@  The top left labels displays the name of an object and the respective age determinated by the PV method.}
\label{fig10}
\end{figure}  

\begin{table}[!t]\centering
 \setlength{\tabnotewidth}{\columnwidth}
 \tablecols{5}
 % Stretch the space between table columns 
 \setlength{\tabcolsep}{0.3\tabcolsep}
 \caption{ Ages ($Age_{PV}$) determined by the PV method for a given evolutionary track mass($M_{Track}$) for the components of the four multiple systems.} 
 \begin{tabular}{lcccc}
  \toprule
Object &$M^{1}_{Track}$&$M_{PV}$&$Age^{2}$(dex)&$Age_{PV}\pm$0.12\\\midrule
GG Tau Aa&0.76             &0.80    & 6.0$\pm$0.35& 6.0\\
GG Tau Ab&0.68             &0.60    &6.03$\pm$0.25&6.25\\
GG Tau Ba&0.12             &0.10    &6.26$\pm$0.39&5.75\\
GG Tau Bb&0.042            &0.10    &$<$6.0$\pm$0.3 &6.75\\
FO Tau A &0.60             &0.40    &6.27$\pm$0.25&6.75\\
FO Tau B &0.59             &0.40    &6.30$\pm$0.25 &6.0\\
FS Tau A &0.53             &0.80    &7.89$\pm$0.3  &6.25\\
FS Tau B &0.20             &0.20    &7.23$\pm$0.38 &6.25\\
V773 Tau AB& 2.6          &2.5     &6.74$\pm$0.17 &5.5 \\    
V773 Tau C&0.7             & 0.6    & $<$6.0$\pm$0.3  &6.0 \\\bottomrule
\multicolumn{5}{l}{1:\citet{bar98}}\\
\multicolumn{5}{l}{2:\citet{whi01}}\\
\end{tabular}
\end{table}

\section{Conclusion}
We explored the effects of excess emission from disks on the JHK-NIR colors on C-C and C-M diagrams.\@  After analizing reprocessing and accretion disk models,  a pattern in J-H vs. H-K and K vs. J-K diagrams was found that gave origin to a set of principal vectors (PV), originated by the IR excess produced by a disk  and  the interstellar extinction.\@ 
We applied a space transformation to verify the possibility of determining the mass of an object at a given age.\@ 

To test our PV method we applied it to a sample of 14 Taurus-Auriga PMS objects, first without resolving the components if they were multiple, with the purpose to determining the object's mass.\@ We found a fair agreement for 12 cases when we used a coeval asumption.\@ For the other 2 cases, in particular for DF Tau, we showed that an agreement with the dynamical mass is reached for a different age, as it had been already pointed out by other authors.\@    

We also applied the method to 4 multiple systems and determined the mass of their individual components, getting a better than 1.5$\sigma$ discrepancy for most of the components when we compared with evolutionary track mass determinations.\@ 

Thus, we have shown that this method works for a particular kind of PMS objects, that is class II classical T Tauri stars.\@ The method still has to be tested on youger (class I) or older systems.\@  Currently we can apply the PV method only to  a mass range of 0.1 to 2.5M$_\odot$.\@ Evolutionary tracks for a wider range of masses would be useful to expand the lower mass limit of the method.\@ For the upper mass limit, we have the additional problem that  the disk vector shows a steeper slope in the C-C D.\@ Accretion models will be needed to further investigate if an apropiate set of vectors for higher stellar masses  can be found.\@ Also, the method needs to be tested in regions with abnormal extinction $A_V > 25$ or R higher than normal, where an appropriate extinction vector would has to be calculated.\@ For the small fraction of stars that would be viewed nearly edge on, gray extinction will appear and the method will fail.\@ However these cases should be sorted out by their unusually larger extinction.\@

Variations in the calculated mass for distance and age uncertainties have been also evaluated.\@ We considered reasonable variations in distance ($7\%$) and age ($22\%$) to the objects in our Taurus sample and found a variation in the $M_{PV}$ masss of 20$\%$.\@ This value is similar to the 20$\%$ residual error of the PV method with respect to dynamical masses, sugesting that most differences can be explained as an age and distance dispersion effect.\@  

Although we mainly used PS99 tracks, we also considered different evolutionary models.\@ The results differed from 20$\%$ to 45$\%$ with most  around 22$\%$ with respect to dynamical masses.\@ The worst comparision is with S00 overshoting models, that have the effect of significant accreted mass on some objects.\@  The better agreement is with PS99 models.\@
 
As a result of this first analysis, we conclude that the PV method is a promising tool to obtain the masses of low to intermediate PMS stars from JHK photometry.\@

For a future work, the method needs to be tested with larger sets of data.\@ Either with known dynamical masses, althougth the advantage of same age and distance would be compromised, or exploring its performance with other clusters with known IMF.\@

\acknowledgments{T.A.L.Ch. was partially supported by CONACyT project No.36574-E.}

\end{document}